\documentclass[journal,transmag]{IEEEtran}
\usepackage[utf8]{inputenc}
\usepackage{graphicx, epsf,multirow}
\usepackage{color,hyperref}
\usepackage[top=0.7in, left=0.65in]{geometry}
\usepackage{pgfplots}
\usepackage{tikz}
\usepackage{siunitx}
\usepackage{circuitikz}

\usepackage{amsmath, amsfonts, bm, amssymb,amscd}

\newcommand{\norm}[1]{\left\lVert#1\right\rVert}
\newcommand \s {{\mathrm s}}
\newcommand \dd {{\mathrm d}}
\newcommand \ddt {\frac{\dd}{\dd t}}
\DeclareBoldMathCommand \cA{{\cal A}}
\DeclareBoldMathCommand \cB{{\cal B}}
\newcommand \cP {\mathbf{p}}
\DeclareBoldMathCommand \cQ{{\cal Q}}
\DeclareBoldMathCommand \cC{{\cal C}}
\newcommand \cW {\mathbf{w}}
\DeclareBoldMathCommand \cI{{\cal I}}
\newcommand \bfw {\mathbf{w}}
\newcommand \bfc {\mathbf{c}}
\newcommand \bff {\mathbf{f}}
\newcommand \bfx {\mathbf{x}}
\newcommand \bfA {\mathbf{A}}
\newcommand \bfB {\mathbf{B}}
\newcommand \bfxhat {\mathbf{\widehat{x}}}
\newcommand \bfchat {\mathbf{\widehat{c}}}

\newcommand \chat {{\widehat{c}}}
\newcommand \Np {N_{\mathrm p}}
\newcommand \Ns {N_{\mathrm s}}
\newcommand \Nsp {N_{\mathrm s}(\Np+1)}
\newcommand \Ts {T_{\mathrm s}}
\newcommand \fs {f_\mathrm{s}}
\newcommand \iL {i_\mathrm{L}}

\newcommand \Ltwo {{\mathrm{L}^2}}
\newcommand{\bfu}{\bar{ \mathbf{w} }}

\pgfplotsset{
  compat=1.13,
  tick label style={font=\footnotesize},
  label style={font=\footnotesize},
  legend style={font=\footnotesize},
}

\usetikzlibrary{external}
\makeatletter
\def\ps@IEEEtitlepagestyle{
  \def\@oddfoot{\mycopyrightnotice}
  \def\@evenfoot{}
}
\def\mycopyrightnotice{
  {\footnotesize
    \begin{minipage}{\textwidth}
      \vspace{3em}
      \fbox{\parbox{\textwidth}{
          \copyright~2017 IEEE. Personal use of this material is permitted. Permission from IEEE must be obtained for all other uses, in any current or future media, including reprinting/republishing this material for advertising or promotional purposes, creating new collective works, for resale or redistribution to servers or lists, or reuse of any copyrighted component of this work in other works.}}
    \end{minipage}
  }
}

\begin{document}
\title{Solving nonlinear circuits with pulsed excitation by multirate partial differential equations}

\author{%
	\IEEEauthorblockN{%
		Andreas Pels\IEEEauthorrefmark{1},
		Johan Gyselinck\IEEEauthorrefmark{2}, 
		Ruth V. Sabariego\IEEEauthorrefmark{3}, and
		Sebastian Schöps\IEEEauthorrefmark{1}
		}
		\IEEEauthorblockA{\IEEEauthorrefmark{1}Graduate School of Computational Engineering and Institut für Theorie Elektromagnetischer Felder,\\Technische Universität Darmstadt, Germany}
		\IEEEauthorblockA{\IEEEauthorrefmark{2}BEAMS Department, Université libre de Bruxelles, Belgium}
		\IEEEauthorblockA{\IEEEauthorrefmark{3}Department of Electrical Engineering, EnergyVille, KU Leuven, Belgium}
		\thanks{Corresponding author: A. Pels (e-mail: pels@gsc.tu-darmstadt.de).}%
}

\IEEEtitleabstractindextext{%
\begin{abstract}
  In this paper the concept of Multirate Partial Differential Equations (MPDEs) is applied to obtain an efficient solution for nonlinear low-frequency electrical circuits with pulsed excitation. The MPDEs are solved by a Galerkin approach and a conventional time discretization. Nonlinearities are efficiently accounted for by neglecting the high-frequency components (ripples) of the state variables and using only their envelope for the evaluation. It is shown that the impact of this approximation on the solution becomes increasingly negligible for rising frequency and leads to significant performance gains.
\end{abstract}

\begin{IEEEkeywords}
  Finite element analysis, Nonlinear circuits, Numerical simulation, Multirate partial differential equations.
\end{IEEEkeywords}}

\maketitle
\IEEEdisplaynontitleabstractindextext
\IEEEpeerreviewmaketitle

\section{Introduction}
\begin{figure}[t]
  \centering
  \includegraphics{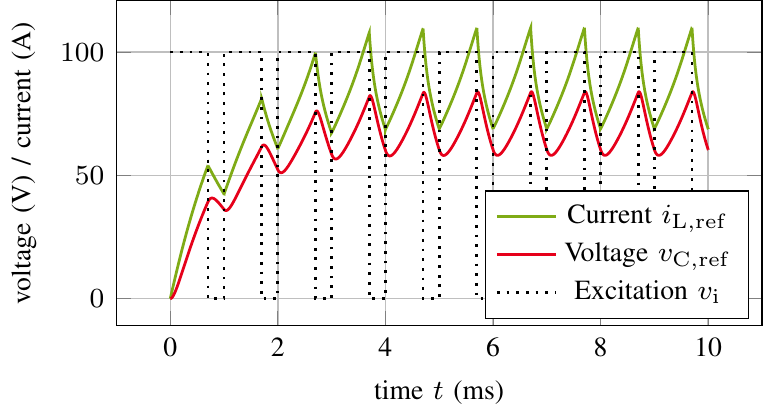}
  \vspace{-1em}
  \caption{Solution of the buck converter shown in Fig. \ref{fig:buckConverter} at $f_\s=1000\,\mathrm{Hz}$. It consists of a slowly varying envelope and fast ripples. The switching cycle of the pulsed input voltage is $\Ts=1\,$ms, the duty cycle $D=0.7$.}
  \label{fig:buckConverterSolution}
  \vspace{-1em}
\end{figure}

\IEEEPARstart{M}{ultiscale} and multirate problems occur naturally in many applications from electrical engineering. Classical discretization schemes are often inefficient in these cases and it is preferable to address the dynamics of each scale separately. In time domain the multirate phenomenon is often characterized by the fact that some solution components are active while the majority is latent (e.g. behind a low-pass filter \cite{Schops_2010aa}) or by problems with oscillatory solutions that are composed of multiple frequencies. An example is depicted in Fig.~\ref{fig:buckConverterSolution}, it consists of a fast periodically varying ripple and a slowly varying envelope.

Multirate Partial Differential Equations (MPDEs) have been successfully applied in nonlinear high-frequency applications with largely separated time scales \cite{Brachtendorf_1996aa, Roychowdhury_2001aa}. Various methods have been proposed for the numerical solution of the MPDEs, e.g. harmonic balance \cite{Brachtendorf_1996aa}, shooting methods, classical time stepping \cite{Mei_2005aa} or a combination of both \cite{Bittner_2014aa}. Instead of the Fourier basis functions, one can also use classical nodal basis functions or more sophisticated problem-specific basis functions as for example the pulse width modulation (PWM) basis functions \cite{Gyselinck_2013ab,Pels_2017ad}.

This contribution focuses on improving the efficiency of the numerical solution of nonlinear MPDEs with a high-frequent pulsed excitation. In contrast to prior works, we propose neglecting the influence of high-frequency components on the nonlinearity. Consequently, the nonlinear relation is evaluated using only the envelope of the state variables.

The paper is structured as follows: after this introduction, Section~\ref{sec:multirate} presents the multirate formulation. Section~\ref{sec:numerics} discusses the Galerkin approach in time domain and the extraction of the envelope to evaluate nonlinear behavior. Section~\ref{sec:experiments} gathers numerical results based on the buck converter benchmark example and discusses the accuracy and efficiency of the modified method. Finally, conclusions are drawn.

\section{Multirate formulation}\label{sec:multirate}
\begin{figure}[t]
  \centering
  \includegraphics{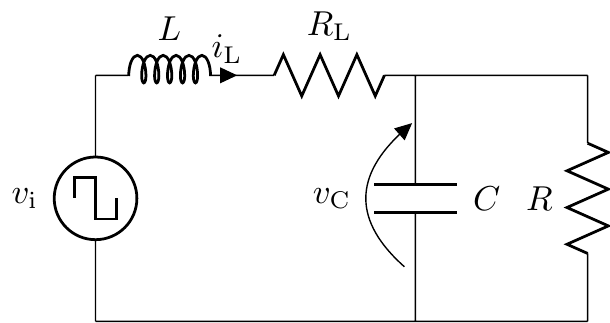}
  \caption{Simplified buck converter.}
  \label{fig:buckConverter}
\end{figure}

Spatial discretization of low-frequency field formulations as, e.g. electro- or magneto-quasi-statics \cite{Haus_1989aa}, or network models of power converter circuits as, e.g. the buck converter in Fig.~\ref{fig:buckConverter} lead to (nonlinear) systems of ordinary or differential algebraic equations of the form
\begin{equation}
  \bfA\bigl(\bfx(t)\bigr) \ddt \bfx(t) + \bfB\bigl(\bfx(t)\bigr) \bfx(t) = \bfc(t)
  \label{equ:DAESystem}
\end{equation}
with an initial condition $\bfx(t_0)=\bfx_0$, where $\bfx(t)\in\mathbb{R}^{\Ns}$ is the vector of $\Ns$ state variables, $\bfA(\bfx), \bfB(\bfx) \in \mathbb{R}^{\Ns\times\Ns}$ are matrices that may depend on the solution and $\bfc(t)\in\mathbb{R}^{\Ns}$ is the excitation vector.

The system of equations~\eqref{equ:DAESystem} is hereafter rewritten as MPDEs \cite{Brachtendorf_1996aa,Roychowdhury_2001aa,Pels_2017ad} in terms of the two time scales $t_1$ and $t_2$
\begin{equation}
  \bfA(\bfxhat) \, \left(\frac{\partial \bfxhat}{\partial t_1} + \frac{\partial \bfxhat}{\partial t_2}\right) + \bfB(\bfxhat) \, \bfxhat(t_1, t_2) = \bfchat(t_1, t_2) \, ,
  \label{equ:MPDESystem}
\end{equation}
where $\bfxhat(t_1, t_2)$ and $\bfchat(t_1, t_2)$ are the multivariate forms of $\bfx(t)$ and $\bfc(t)$. 
If $\bfchat(t_1, t_2)$ fulfills the relation $\bfchat(t,t)=\bfc(t)$, the solution of the original problem can be extracted from the solution $\bfxhat(t_1,t_2)$ of the MPDEs by $\bfx(t)=\bfxhat(t,t)$ \cite{Brachtendorf_1996aa}. 

Let $t_1$ denote the slow time scale and $t_2$ the fast time scale, which is furthermore assumed to be periodic. Without limiting the generality of the approach, we choose $\bfchat(t_1,t_2):=\bfc(t_2)$ such that the right-hand-side only depends on the fast time scale \cite{Pels_2017ad}.

\section{Numerical method}\label{sec:numerics}
System \eqref{equ:MPDESystem} can be numerically solved either using a Galerkin framework, shooting methods, classical time stepping or a combination. Here, we propose to use a variational setting, i.e. a Galerkin approach, for the fast time scale and a conventional time stepping for the slowly varying envelope.

\subsection{Galerkin in time domain}
We represent the solution by an expansion of $\Np+1$ suitable basis functions $p_k(\tau(t_2))$ and coefficients $w_{j,k}(t_1)$. The approximated state variables $\bfxhat_j(t_1,t_2)$ can be written as the series
\begin{equation}
  \bfxhat_j(t_1,t_2)=\sum\limits_{k=0}^{\Np} p_k(\tau(t_2)) w_{j,k}(t_1),
  \label{equ:solExp}
\end{equation}
with $\tau(t_2)=\frac{t_2}{\Ts}\text{ mod }1$, where $\Ts$ is the switching cycle of the excitation and mod denotes the modulo operation. Since we deal with pulsed right-hand-sides, we choose the PWM basis functions of \cite{Pels_2017ad}, although constructed for linear problems \cite{Gyselinck_2013ab}. The zero-th basis function is constant $p_0(\tau)=1$ and the corresponding coefficient $w_{j,0}$ defines the envelope of the $j$-th solution component.
The first basis function is defined by
\begin{equation*}
  \renewcommand*{\arraystretch}{1.3}
  \displaystyle
  p_1(\tau) \ = \ \left\{  
  \begin{array}{ll} 
    \sqrt{3} \ {2\tau-D \over D} \quad & \text{if~} 0 \leq \tau \leq D \\ 
    \sqrt{3} \ {1+D-2\tau \over 1-D}  \quad & \text{if~} D \leq \tau \leq 1    \end{array}
  \right.,
\end{equation*}
where $D$ is a free parameter, which can be chosen according to the duty cycle of the PWM, i.e. between 0 and 1.
The basis functions of higher degree $p_k(\tau)$, $2\leq k \leq \Np$ are obtained recursively by integrating $p_{k-1}(\tau)$ and orthonormalizing with respect to the $\Ltwo(0,1)$ scalar product, \cite{Gyselinck_2013ab,Pels_2017ad}.

Finally, a Galerkin approach is applied on the interval of one period, which yields 
\begin{equation}
  \begin{aligned}
    \int\limits_{0}^{T_{\s}}
    \biggl( &\bfA(\bfxhat) \, \Bigl(\frac{\partial \bfxhat}{\partial t_1} + \frac{\partial \bfxhat}{\partial t_2}\Bigr) \\
    +& \bfB(\bfxhat) \, \bfxhat 
    -\bfchat \biggr) \, p_l(\tau(t_2)) \, \dd t_2 = 0,
    \label{equ:gal}
  \end{aligned}
\end{equation}
for all $l=0,\dots,\Np$. The integration with respect to $t_2$ leads to a system of differential (algebraic) equations in $t_1$, which can be solved by conventional time integration. 
	
\begin{figure}
  \centering
    \includegraphics{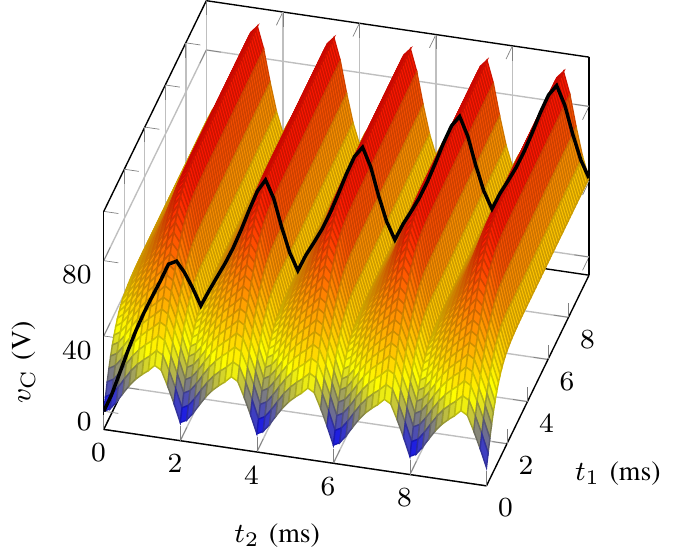}
    \vspace{-0.5em}
    \caption{Illustration of the multirate solution for $\fs=500\,\mathrm{Hz}$. The solution of the original system of differential equations is denoted in black.}
    \label{fig:MPDE3DSolution}
    \vspace{-1em}
\end{figure}

\subsection{Treatment of nonlinearity}
During time integration of \eqref{equ:gal}, the integrals have to be evaluated every time step due to their nonlinear dependency on the solution. This may lead to an unnecessary increase of the computational effort as the ripple components are often small in comparison with the magnitude of the envelope. Consequently, one can save computational time by ignoring the ripple components of the solution and only using its envelope $\bfu$ for the evaluation of the nonlinearity.

As mentioned before, the envelope is stored in the vector of coefficients $\bfw(t_1)$ given by
\begin{equation}
  \bfw=[w_{1,0}, \dots, w_{1,\Np}, w_{2,0},\dots, w_{\Ns,\Np}]^\top. 
\end{equation}
Let us abstractly define a function $\bff$, which extracts the envelope $\bfu(t_1)$ from $\bfw(t_1)$, i.e.
 \begin{equation}
   \bfu(t_1)=\bff(\bfw(t_1)), 
 \end{equation}
 which is in the case of PWM basis functions the zero-th components, i.e. 
 \begin{equation}
   \bfu=\left[
     w_{1,0},
     w_{2,0},
     \dots,
     w_{\Ns,0}
     \right]^\top.
 \end{equation}
Therefore the matrices $\bfA$, $\bfB$ only depend on $t_1$ and the evaluation of \eqref{equ:gal} simplifies significantly. Equation \eqref{equ:gal} becomes
\begin{equation}
  \begin{aligned}
    \int\limits_{0}^{T_{\s}} \biggl( &\bfA(\bfu) \, \Bigl(\frac{\partial \bfxhat}{\partial t_1} + \frac{\partial \bfxhat}{\partial t_2}\Bigr)\\
    + &\bfB(\bfu) \, \bfxhat 
    -\bfchat \biggr) \, p_l(\tau(t_2)) \, \dd t_2 = 0,
    \label{equ:galConst}
  \end{aligned}
\end{equation}
for all $l=0,\dots,\Np$, where the matrices $\bfA$ and $\bfB$ are independent of $t_2$. Introducing
\begin{align}
  \cI  \ &= \ \Ts \int_0^1  \cP(\tau) \, \cP^\top \!(\tau)  \,\dd \tau \, ,\\
  \cQ \ &= - \ \int_0^1  \frac{\dd \cP}{\dd \tau } \, \cP^\top(\tau) \, \dd \tau  \, ,
\end{align}
in equation \eqref{equ:galConst}, we get
\begin{equation}
  \cA(\bfw(t_1))\, \frac{\dd  \cW}{\dd t_1} + \cB(\bfw(t_1)) \, \cW(t_1)  \ = \ \cC(t_1) \, ,
  \label{equ:ReducedMPDESystem}
\end{equation}
where the matrices are given by
\begin{align}
  &\cA(\bfw)=\bfA(\bff(\bfw))\otimes\cI,\\
  &\cB(\bfw)=\bfB(\bff(\bfw))\otimes\cI+ \bfA(\bff(\bfw))\otimes\cQ,
\intertext{and $\otimes$ denotes the Kronecker product. The right-hand-side vector is given by}
  &\cC(t_1) = \int_{0}^{T_\s}
  \left[\begin{array}{cccc}
    \chat_{1}(t_1,t_2)  \,\cP(\tau(t_2))  \\
    \vdots \\
    \chat_{\Ns}(t_1,t_2)  \,\cP(\tau(t_2))  \\ 
  \end{array}\right]
  \dd t_2 \,.
  \label{equ:integralC}
\end{align}
\begin{figure}[t]
  \centering
  \includegraphics{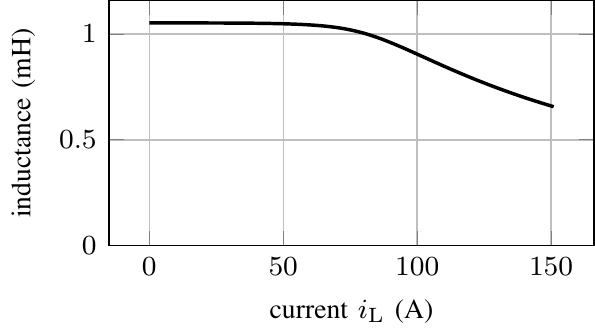}
  \vspace{-0.5em}
  \caption{Inductance of nonlinear coil versus current $\iL$ through the coil.}
  \label{fig:nonlinCoil}
\end{figure}
Eventually, system \eqref{equ:ReducedMPDESystem} can be more efficiently solved by conventional time discretization in the sense that larger time steps can be used than for the original problem \eqref{equ:DAESystem}. However, drawbacks are the approximation of the nonlinearity and the increased size of the matrices, i.e. $\cA(\bfw)$, $\cB(\bfw) \in \mathbb{R}^{\Nsp\times\Nsp}$ and $\cC(t)\in \mathbb{R}^{\Nsp}$, which is a similar tradeoff as in harmonic balance \cite{Brachtendorf_1996aa}.

\section{Numerical Results}\label{sec:experiments}
The numerical tests are performed on the simplified buck converter model \cite{Gyselinck_2013ab} using a nonlinear coil, whose characteristic is shown in Fig.~\ref{fig:nonlinCoil}. The code is implemented in GNU Octave \cite{Eaton_2015aa}; for time integration the high-order implicit Runge-Kutta method Radau5 from odepkg is used, \cite{Hairer_1999aa,Treichl_2015aa}. As basis functions $p_k(\tau)$ we choose the problem-specific PWM basis functions introduced earlier.

The reference solution to which all results are compared is calculated directly by solving~\eqref{equ:DAESystem} with a very accurate time discretization ($\text{tol}=10^{-12}$).  
The buck converter is operated in the range of frequencies from $500\,\mathrm{Hz}$ to $100\,\mathrm{kHz}$. To determine the accuracy of the MPDE approach the relative $\Ltwo$-error of the buck converter voltage
\begin{equation}
  \epsilon=\frac{\norm{v_{\mathrm{C,ref}}-{v_{\mathrm{C}}}}_{\Ltwo(\Omega)}}{ \norm{v_{\mathrm{C,ref}}}_{\Ltwo(\Omega)}}
\end{equation}
with respect to $t\in \Omega=[0,10]\,\mathrm{ms}$ is approximated by numerical quadrature. This calculation is performed for each frequency $f_\s= \frac{1}{\Ts}$. 

\begin{figure}[t]
  \centering
  \includegraphics{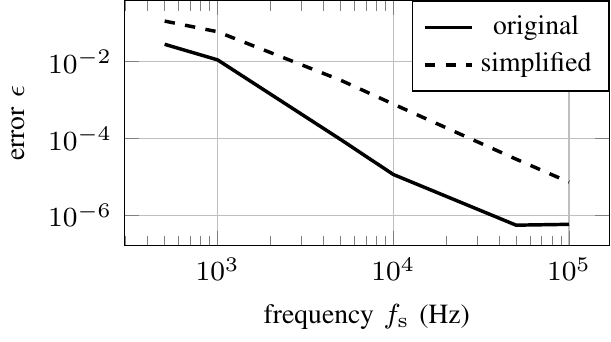}
  \vspace{-0.5em}
  \caption{Error $\epsilon$ of the original and simplified MPDE approach ($\Np=4$) versus frequency $f_\s$.}
  \label{fig:convergencePWMBF}
  \vspace{-1em}
\end{figure}

\begin{figure}[t]
  \centering
  \includegraphics{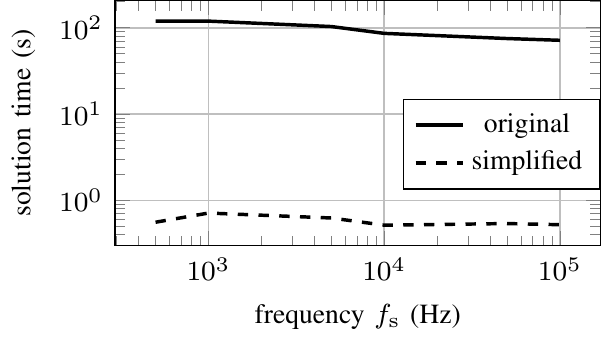}
  \vspace{-0.5em}
  \caption{Computational time for the original MPDE formulation and the one with simplified evaluation of nonlinearity ($\Np=4$).}
  \label{fig:timePWMBF}
  \vspace{-1em}
\end{figure}

\begin{figure}
  \includegraphics{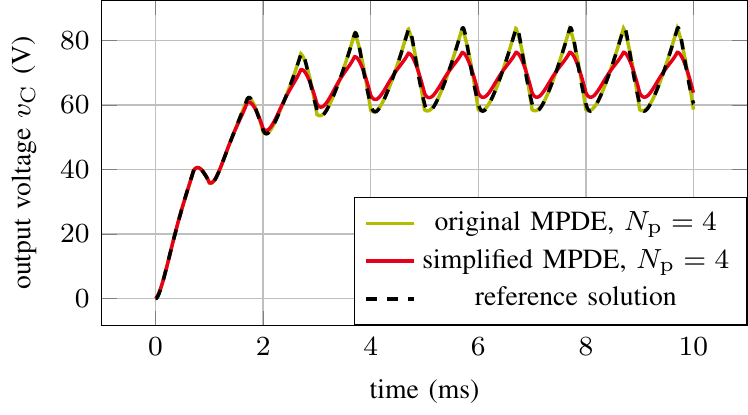}
  \vspace{-0.5em}
  \caption{Comparison of solutions for $\fs=1000\,$Hz.}
  \label{fig:comparisonSolutions}
\end{figure}

The MPDE solution is expanded using $\Np=4$ basis functions and \eqref{equ:ReducedMPDESystem} is solved using a tolerance of $\text{tol}=10^{-6}$. The result is exemplary depicted in Fig. \ref{fig:MPDE3DSolution}. 
Fig. \ref{fig:convergencePWMBF} shows the error of the approach without (denoted as ``original approach'') and with the simplified evaluation of the nonlinearity \eqref{equ:galConst} (denoted as ``simplified approach'') with respect to the frequency. Without the simplification, the integrals in \eqref{equ:gal} are evaluated using Gauss-Kronrod quadrature and lead to an accuracy of $\epsilon<10^{-4}$ for frequencies $\fs>10\,$kHz. As expected, the higher the frequency, the higher the accuracy of the method since the magnitude of the ripples in relation to the envelope decreases. 
The simplified approach introduces an additional error due to the approximation of the integrals. However, the accuracy for $f_\mathrm{s}>10\,$kHz, i.e., $\epsilon<10^{-3}$, is still sufficient for most applications. Fig. \ref{fig:comparisonSolutions} shows the solutions of reference, simplified and original MPDE approach versus time at a low frequency $\fs=1000\,$Hz. Here, the error committed in the simplified approach is clearly distinguishable. 

Table~\ref{tab:timeDiscretizationDuration} shows the speedup (in terms of time for solving the equation systems) of the simplified MPDE approach compared to a conventional time discretization at the same accuracy. For higher frequency, the conventional time discretization of~\eqref{equ:DAESystem} becomes more and more inefficient as a higher number of ripples has to be resolved. The MPDE approach on the contrary resolves the ripples with the Galerkin approach so that the time discretization resolves only the envelope. This leads to a solution time almost independent of the frequency, i.e., approximately $1\,$s for the simplified and $100\,$s for the original approach, see Fig.~\ref{fig:timePWMBF}. The higher solution time of the original approach is a result of the evaluation of the integrals in each time step. Thus the speedup of the original approach is much lower compared to the simplified approach.

\section{Conclusion}
The MPDE approach is applied to a nonlinear low-frequency example with pulsed excitation. The solution is obtained by a Galerkin approach and time discretization. To evaluate the nonlinearity the ripple components due to the pulsed excitation are neglected and only the envelope is used. The accuracy of the proposed method rises with increasing excitation frequency and the method offers a considerable speedup compared to conventional time discretization with the same accuracy. 

\section*{Acknowledgement}
  This work is supported  by the `Excellence Initiative' of German Federal and State Governments and the Graduate School CE at TU Darmstadt and in part by the Walloon Region of Belgium (WBGreen FEDO, grant RW-1217703).

\begin{table}
  \caption{Speedup of MPDE approach ($\Np=4$) compared to conventional time discretization for different frequencies.}
  \label{tab:timeDiscretizationDuration}
  \centering
    \begin{tabular}{|c|c|c|}
      \hline 
      $f_\s$ (kHz) & approx. speedup & approx. error\\ \hline
      10 & 60 & $8\cdot 10^{-4}$ \\ \hline
      50 & 400 & $3\cdot 10^{-5}$ \\ \hline
      100 & 1000 & $7\cdot 10^{-6}$\\ \hline
    \end{tabular}
\end{table}

\bibliography{abbrv,english,library}
\bibliographystyle{IEEEtran}

\end{document}